\documentstyle[twocolumn,prb,aps]{revtex}
\begin{document}

\title{Superconductivity with \textit{s} and \textit{p} symmetries in an extended
Hubbard model with correlated hopping
\thanks{Dedicated to Prof. Dr. J. Zittartz on the occasion of his
$60^{th}$ birthday}}
\author{A.A.Aligia$^{1}$, E. Gagliano$^{1}$, Liliana Arrachea$^{2}$ and K.
Hallberg$^{1}$}

\address{$^1$ Centro At\'omico Bariloche and Instituto Balseiro, Comisi\'on
Nacional de Energ{\'\i}a At\'omica, 8400 Bariloche, Argentina \\  
$^2$ Max-Planck-Institut f\"{u}r Physik komplexer Systeme, N\"othnitzer Str. 38,
01187 Dresden, Germany }

\maketitle

\begin{abstract} 
We consider a generalized Hubbard model with on-site and
nearest-neighbour repulsions $U$ and $V$ respectively, and nearest-neighbour
hopping for spin up (down) which depends on the total occupation $n_{b}$ of
spin down (up) electrons on both sites involved. The hopping parameters are $%
t_{AA}$, $t_{AB}$ and $t_{BB}$ for $n_{b}=0,1,2$ respectively. We briefly
summarize results which support that the model exhibits \textit{s}-wave
superconductivity for certain parameters and extend them by studying the
Berry phases. Using a generalized Hartree-Fock(HF) BCS decoupling of the two
and three-body terms, we obtain that at half filling, for $%
t_{AB}<t_{AA}=t_{BB}$ and sufficiently small $U$ and $V$ the model leads
to triplet \textit{p}-wave superconductivity for a simple cubic
lattice in any dimension. In one dimension, the resulting phase diagram is
compared with that obtained numerically using two quantized Berry phases
(topological numbers) as order parameters. While this novel method supports
the previous results, there are quantitative differences. 
\end{abstract}


\narrowtext


\section{Introduction}

\label{intro}

Since the discovery of high $T_{c}$ much effort has been devoted to the
study of the pairing mechanisms in highly correlated electronic systems. Two
very well studied microscopic models for such systems are the Hubbard and $%
t-J$ models. While the search for superconductivity in the positive-$U$
Hubbard model by numerical methods has failed so far \cite
{lin2,imad,more,dago,zhan}, $t-J$-like models exhibit \textit{d}-wave
superconductivity for certain parameters \cite{rier,heeb,bati,oga,lema}.

Although most of the high $T_{c}$ superconducting materials have a
two-dimensional character, calculations in one dimension (1D) have been very
useful. In one dimension the Hubbard and supersymmetric ($J=2t$) $t-J$
models can be exactly solved using the Bethe ansatz \cite{lw,tj}. It has
been shown that for this particular case of the $t-J$ model, the ground
state consists of bound states with a gapless excitation spectrum, but for $%
J<2t$ bound states exist for large enough densities \cite{carlos}.

Other strongly correlated models that introduce correlated hopping
interactions are subject of current research [14-51]. In
1989, Hirsch \cite{2} proposed a model for the description of oxide
superconductors by considering the holes in a nearly filled band as charge
carriers. The Hamiltonian in standard notation reads:

\begin{eqnarray}
\label{eqn:1}
H_H=-\sum_{j=1}^L \sum_{\sigma=\pm 1} (c^{\dagger}_{j,\sigma} 
c_{j+1,\sigma} + c^{\dagger}_{j+1,\sigma} c_{j,\sigma})\times  \nonumber \\
\left ( 1-\frac{\Delta t}{2}(n_{j,-\sigma}+
n_{j+1,-\sigma})\right )
\end{eqnarray}

In contrast to the Hubbard model, the hopping amplitude for single-particle
hopping to a nearest-neighbour site depends on the occupancy of one of the
sites involved in the process.

A few years ago, a new integrable model of strongly correlated fermionic
systems for hole superconductivity has been introduced \cite{3,4}. It is a
modified version of Hirsch's model (\ref{eqn:1}) and its Hamiltonian reads:

\begin{eqnarray}
\label{eqn:2}
H_B=-\sum_{j=1}^L \sum_{\sigma=\pm 1} (c^{\dagger}_{j,\sigma} 
c_{j+1,\sigma} + c^{\dagger}_{j+1,\sigma} c_{j,\sigma})\times \nonumber \\
\left ( 1-\Delta tn_{j+(1+\sigma)/2,-\sigma} \right )
\end{eqnarray}

In 1D, it has been solved using the Bethe ansatz in three parameter regimes: 
$0<\Delta t<1$, $\Delta t<0$ \cite{3,4,6} and $\Delta t>1$ \cite{andreas}.
The two first regimes are related by a particle-hole transformation and
present similar behaviour to the Hirsch model (\ref{eqn:1}): there is a good
quantitative agreement between the ground state energies, both models show
gapless charge excitations and spin excitations with a finite gap for all
densities and also dominating superconducting correlations for densities
larger than a critical value \cite{quaisser}. Instead, for the case with $%
\Delta t>1$, both models (\ref{eqn:1} and \ref{eqn:2}) behave quite
differently: while the spin gap in the Hirsch model vanishes and there are
no indications of superconductivity, the gap remains finite and
superconducting correlations are present for small doping in the Bariev
model (\ref{eqn:2}).

An important interaction that has not been considered in the above models is
the Coulomb repulsion. We consider a more complete Hamiltonian that includes
on-site $U$ and nearest-neighbour $V$ interactions between the
particles, and contains a more general kinetic term with different
hopping parameters, depending on the total occupation of the sites
involved:

\begin{eqnarray}
\label{eqn:3}
H&=&-\sum_{<ij>\sigma} \Phi_{ij} (c^{\dagger}_{i,-\sigma}
c_{j,-\sigma}+hc) + U \sum_i n_{i \uparrow}   n_{i \downarrow} + 
\nonumber \\
&~&V \sum_{<ij>} n_i   n_j ;  \nonumber \\
  \Phi_{ij}&=&t_{AA}(1-n_{i\sigma}) (1-n_{j\sigma}) + t_{BB}
  n_{i\sigma}  n_{j\sigma} + \nonumber \\
&~&t_{AB} ( n_{i\sigma} +  n_{j\sigma} -2
  n_{i\sigma}  n_{j\sigma})
\end{eqnarray}

The model (\ref{eqn:1}) corresponds to taking
$t_{AA}+t_{BB}-2t_{AB}=0$, $U=V=0$
and $\Delta t/2=1-t_{AB}/t_{AA}$, and the Hubbard model is obtained when $%
V=0,$ $t_{AA}=t_{BB}=t_{AB}$. The case $V=0$ has been derived as the model
that describes the low energy excitations of intermediate-valence systems 
\cite{foglio} and ``hole'' and cuprate superconductors \cite{marsiglio,simon}%
. The model with $V$ was obtained for cuprate superconductors \cite
{fedro,den} and proposed to describe the benzene molecule \cite{camp,voll}.

For $t_{AB}=V=0$ and $|t_{AA}|=|t_{BB}|$ an exact solution of (\ref{eqn:3}) exists in 1D \cite
{exa1,8,11,afq}, but this case is too peculiar, as discussed later.
Hamiltonian (\ref{eqn:3}) supplemented by a hopping of pairs of the type $%
t_p \sum_{\langle ij \rangle}
 c_{i,\uparrow }^{\dagger }c_{i,\downarrow }^{\dagger }c_{j,\uparrow
}c_{j,\downarrow }$ has also been solved in 1D using Bethe ansatz \cite
{karn,bedu,brac} for $U=-2t_{p}$ and $%
t_{AA}t_{BB}=t_{AB}^{2}$. Its behaviour is similar to that of the Hubbard
model. For negative $U$ the model displays superconductivity.
A more general model including this one and that of Bariev for
particular parameters has also been solved using Bethe ansatz \cite{18}.

 For $t_{AB}=0, |t_{AA}|=|t_{BB}|=t, U, V \neq 0$
the phase diagram of the model (\ref{eqn:3}) has been investigated
previously \cite{voll,ovch,14,aligia} and some exact
results have been found for the half-filled system (density of particles $n=1$):
 a) in any arbitrary lattice
in $D$ dimensions with coordination number $z=2D$ the ground state is a Mott
insulator (MI) with all sites singly occupied if $U>z$ $%
max(V,|t_{AA}|+|t_{BB}|)$;
b) the ground state is a charge density wave (CDW) if $%
V>max(U/z,U/2+|t_{AA}|+|t_{BB}|)/2$ (simple cubic lattices) ; c) 
there is a region with mobile carriers which we call metallic (M) between
these two phases, the M-MI boundary being at $U_{M-MI}=z(|t_{AA}|+|t_{BB}|)$
(for $D=1$ this state is a ``non-conducting metal'' \cite{afq,14}); d) for $%
D=1$ the M-CDW boundary is at $V_{M-CDW}=(U/2+|t_{AA}|+|t_{BB}|)/2$. 

 For $0<t_{AB}<t_{AA}=t_{BB}=t, U, V \neq 0$ the phase diagram
at $n=1$ has been studied 
numerically in 1D and 2D and within mean-field approximation
\cite{aligia,mon,mit}. The high spin degeneracy of the MI phase
for $t_{AB}=0$
is lifted and gives place to a spin density wave (SDW) phase.
There exists a metallic phase (the detailed nature of which will be
discussed later)
for small values of $U$ and $V$, which
shrinks as $t_{AB} \rightarrow t$.
When $t_{AB}=t$ (Hubbard limit), several results indicate that there is no M phase,
except, eventually, on the second-order transition line between the CDW and
SDW phases, that ends at a tricritical point. The position of this
point is, to
date, not well determined and should be around $U\sim 2V\sim 4t$ \cite
{hir2,cann2,cann,four,voit}.

The aim of the present work is to study, by numerical and analytical
methods, possible existence of superconductivity in the model (\ref{eqn:3})
in regions of parameters for which exact results are not available. Special
attention is paid to the occurrence of exotic \textit{p}-wave triplet
superconductivity (TS). In particular, we show strong evidence that the
extension to finite $t_{AB}$ of the above mentioned M phase displays TS.
Some evidence of triplet superconductivity exists in the extended Hubbard
model (Eq.~\ref{eqn:3} with $t_{AA}=t_{AB}=t_{BB}$) very near the line
$U=-2V$ for positive $U$ \cite{voit,lin}.

In Section 2 we briefly review the results which indicate that the model (%
\ref{eqn:3}) exhibits \textit{s}-wave superconductivity (or dominant singlet
superconducting correlations at large distances in 1D) for certain
parameters. At half filling we present evidence of a
superconductor-insulator transition. In Section 3 we explain the HF-BCS
decoupling scheme and apply it to the electron-hole symmetric case $%
t_{AB}<t_{AA}=t_{BB}$ at half filling. We obtain a phase diagram separating
regions in which the stable phase is a CDW, SDW or triplet \textit{p}-wave
superconductor (TS). The latter is the stable one for sufficiently small $U$
and $V$ in a simple cubic lattice in any dimension. In Section 4 we
calculate the phase diagram in 1D by a numerical method recently introduced
by two of us \cite{berry}, which uses topological quantum numbers as order
parameters, and compare with the HF-BCS results. Section 5 contains the
conclusions.

\section{\textit{s}-wave superconductivity}

The model (\ref{eqn:3}) becomes simpler when the three-body part of it
vanishes. The coefficient of terms of the form
$c^{\dagger}_{i,\sigma}c_{j,\sigma}$ 
$n_{i,-\sigma}n_{j,-\sigma}$ is $t_3=2t_{AB}-t_{AA}-t_{BB}$. 

We concentrate first in the results for
$t_3=0$. Using a HF-BCS decoupling, it has been shown that the model leads to
(extended) \textit{s}-wave superconductivity for $V=0$, small enough $U$ and
positive (negative) $t_2=t_{AA}-t_{AB}$ for a more (less) than half-filled
band \cite{mar}. In 1D Japaridze and M\"uller-Hartmann have calculated the
correlation exponent $K_{\rho}$ using continuum limit theory and
bosonization \cite{10}. They obtained that $K_{\rho}>1$ (superconducting
correlations dominate at large distances) if

\begin{equation}
U<U_{c}=8(t_{AB}-t_{AA})\cos (\pi n/2)-6V  \label{eqn:4}
\end{equation}
where $n$ is the number of particles per site.

 The particular case $t_{2}=t_{AA}$  corresponds to the relation
$t_{AA}=-t_{BB},t_{AB}=0$,
and has been exactly solved for $V=0$ \cite{exa1,8,11,afq}.
Superconducting $\eta $%
-paired states with off-diagonal long-range order (ODLRO) are part of the
highly degenerate ground state. However, as a consequence of this
degeneracy, the system does not display anomalous flux quantization (AFQ)
for large rings \cite{afq}. In other words, the Meissner effect is absent.

Since, due to the rather pathological degeneracy of the ground state, the
exact results do not allow to draw definite conclusions for $t_{AB}\neq 0$
while the bosonization results are expected to be valid only for weak
interactions, we have carried out a detailed numerical study of the model
for $V=0$ in rings of 10 and 12 sites \cite{13}. For $t_{AA}=1$, $t_{AB}=2$, 
$t_{BB}=3$, $2/3<n<4/5$ and small $U$, there are clear indications of
binding and AFQ (a tendency towards a periodicity of half a flux quantum in
the energy as a function of flux). The calculation of the correlation
exponent $K_{\rho}$ indicates that superconducting correlations dominate for $%
U<U_c$, where $U_c\sim 9$, 8 or 6.5 for $n=0.4$, 0.5 or 0.6 electrons per
site respectively. These values are nearly 1.4 times larger than the
corresponding continuum limit results: 6.46, 5.66 and 4.70 respectively,
according to Eq.~(\ref{eqn:4}). For $n=0.8$, $t_{AA}=1$, $t_{AB}=1.5$ and $%
t_{BB}=2$ we obtain $U_c=1.6$, while Eq.~(\ref{eqn:4}) gives $U_c=1.24$.

The qualitative agreement between Eq.~(\ref{eqn:4}) and our numerical
results is lost at half filling. While the continuum limit theory predicts
that the system is an insulator for all positive values of $U$ \cite{10}, we
find evidence of a superconductor-insulator transition as a function of $U$
when $t_{3}=2t_{AB}-t_{AA}-t_{BB}=0$ and any sign of $t_{2}=t_{AA}-t_{AB}$
(the sign can be changed using the symmetry properties of the model \cite
{afq,13}). This agrees with previous numerical work of $K_{\rho }$ \cite
{13,airo} , a BCS calculation \cite{airo} and recent results using
slave bosons \cite{bulka}. For small positive values of $U$
we find in rings of 10 sites that the energy as a function of flux $E(\phi )$
has a form that suggests AFQ for $t_{AA}=1$, $t_{AB}=0.5$ and $t_{BB}=0$.
Furthermore, using topological quantum numbers \cite{berry}
(as explained briefly in Section 4), we detect a transition from charge 
($\gamma_c$) and spin ($\gamma_s$)
Berry phases $ (\gamma_c,\gamma_s)=(0,0)$ to $(\pi ,\pi )$ (corresponding to the SDW phase) as $U$
increases. We point out that the topological values $(0,0)$ are also
obtained for the negative-$U$ Hubbard model which displays singlet \textit{s}%
-wave superconductivity. The value of $U_{c}$ separating both phases is
small. For 10 sites we obtain $U_{c}=0.075$. From the size-dependence we
estimate that this actually corresponds to a lower bound. From the results
of $K_{\rho }$ of Ref.~\cite{13} and using symmetry arguments we
estimate (remember $n=1$) $%
U_{c}=0.5$ for $t_{AA}=1$, $t_{AB}=3/5$ and $t_{BB}=1/5$ and $U_{c}=0.3$ for 
$t_{AA}=1$, $t_{AB}=2/3$ and $t_{BB}=1/3$, but careful finite-size scaling is
necessary to give accurate values of $U_{c}$.

While the values of $K_{\rho }$ alone cannot distinguish between singlet 
\textit{s}-wave (even) and triplet \textit{p}-wave (odd) superconducting
states, the BCS results and the Berry phases (0,0) are indicative of the
former. A demonstration of the \textit{s}-wave character was provided by the
results of stochastic diagonalization by Michielsen and De Raedt \cite{mich}
which showed the presence of singlet-singlet quasi ODLRO in 1D for $t_{AA}=1$%
, $t_{AB}=1.4$ and $t_{BB}=1.8$, $n=1.5$ and $U<1$, and also for $%
t_{AA}=t_{AB}=t_{BB}=1$, $U=-4$ and $n=1.5$ (negative-$U$ Hubbard model).

We discuss now the effect of  
the three-body term $t_{3}=2t_{AB}-t_{AA}-t_{BB}$.
When $t_{AA}=t_{BB}$ the model is electron-hole symmetric
for bipartite lattices \cite{13}.
For the electron-hole symmetric case  with $t_3 > 0$,
 the mean-field HF-BCS
decoupling of the two- and three-body terms leads to singlet {\it s}-wave
(and also {\it d}-wave in 2D) superconducting solutions \cite{har}. However, a
positive $t_{3}$ also favours CDW and SDW instabilities \cite{mon}. In 2D,
the detailed (mean-field)
calculation of the phase diagram for $V=0$ shows that the SDW
is stable for positive $U$ near half filling, while extended {\it s}%
-wave superconductivity is present in the ground state for small $U$ or
large doping. The {\it d}-wave paramagnetic solution has always larger
energy than the other two \cite{har}.
The  electron-hole symmetric  case with $t_3 < 0$ is analyzed in the next section.

\section{\textit{p}-wave superconductivity}

In this Section we discus the phase diagram of the model for the
electron-hole symmetric case $t_{AB}<t_{AA}=t_{BB}$ and half filling, in
simple cubic lattices in arbitrary dimensions using HF-BCS. This decoupling
leads naturally to \textit{p}-wave superconductivity for small $U$ and $V$.
Although a real proof of the existence of this phase is lacking, further
arguments given in the next Section support its existence in 1D. For $%
t_{AB}=0$, exact results have shown that for small $U$ and $V$, the CDW and
SDW phases become unstable because the system lowers its energy taking
advantage of the kinetic energy terms $t_{AA}$ and $t_{BB}$, which are
inactive in the CDW and SDW phases \cite{voll,ovch,14}. As mentioned in the
Introduction, a phase diagram valid for several lattices in arbitrary
dimensions was constructed \cite{14}. For $t_{AB}=V=0$ , an exact solution
exists in 1D \cite{exa1,8,11,afq}, but these exact results were unable to
identify the nature of the third phase for finite $t_{AB}$.
The numerical and mean-field results \cite{aligia,mon} 
show the presence of
mobile carriers and a non-vanishing Drude weight within the M phase
 in 1D and 2D, suggesting that the system is a Luttinger liquid in 1D
and metallic in 2D for small $U$ and $V$. However, a suggestion that this
phase has dominant triplet superconducting correlations at large distances
was given only recently \cite{berry}.

Note that if in the three-body part of the correlated hopping (see Eq.~\ref
{eqn:3}), the operator $c^{\dagger}_{i,\sigma}c_{j,\sigma}$ is replaced by
its expectation value $\tau$(assumed for this argument independent of spin
and nearest-neighbour pair, as in the non-interacting case), this term takes
the form $t_3 \tau \sum_{\langle ij \rangle} n_{i,-\sigma} n_{j,-\sigma}$, with $%
t_3=2t_{AB}-t_{AA}-t_{BB}<0$. Thus, the three-body term reduces to an
attraction of nearest-neighbour electrons \textit{with the same spin} and
triplet pairing is a natural consequence of it.

The HF-BCS decoupling is the most convenient mean-field approximation to
reduce the many-body terms of the Hamiltonian to one-body terms. It is a
generalization of the procedure used by Foglio and Falicov for the normal
case \cite{foglio}. A more direct way of obtaining the HF-BCS
Hamiltonian ($H_{HFBCS}$) is
to define the vacuum as the (unknown for the moment) Slater determinant that
is the ground state of $H_{HFBCS}$. Then one should normal order the exact Hamiltonian with
respect to this vacuum using Wick's theorem \cite{blai,lore}. Clearly the
contractions that appear using this theorem are the HF-BCS expectation
values. The exact Hamiltonian takes the form of the HF-BCS ground-state
energy plus normal ordered one- and many-body terms. Neglecting the latter,
one obtains the HF-BCS Hamiltonian.

Neglecting also for simplicity spin-flip expectation values of the form $%
\langle c_{i,\sigma }^{\dagger }c_{j,-\sigma }\rangle $
(related to spiral spin
structures), we obtain $H_{HFBCS}$ using the following approximations for the two- and
three-body terms of $H$:

\begin{eqnarray}
\label{de1}
n_{i\uparrow }n_{i\downarrow } &\simeq &\langle n_{i\uparrow }\rangle
n_{i\downarrow }+n_{i\uparrow }\langle n_{i\downarrow }\rangle -\langle
n_{i\uparrow }\rangle \langle n_{i\downarrow }\rangle  \nonumber \\
&&+(c_{i\uparrow }^{\dagger }c_{i\downarrow }^{\dagger }\langle
c_{i\downarrow }^{{}}c_{i\uparrow }^{{}}\rangle +h.c.)-|\langle
c_{i\downarrow }^{{}}c_{i\uparrow }^{{}}\rangle |^{2},  
\end{eqnarray}

\begin{eqnarray}
\label{de2}
(n_{i\uparrow}+n_{i\downarrow })&&(n_{j\uparrow}+n_{j\downarrow }) \simeq
\langle n_{i\uparrow }+n_{i\downarrow }\rangle (n_{j\uparrow
}+n_{j\downarrow }) \nonumber \\
&&+(n_{i\uparrow}+n_{i\downarrow })\langle n_{j\uparrow
}+n_{j\downarrow }\rangle  
-\langle n_{i\uparrow }+n_{i\downarrow }\rangle \langle n_{j\uparrow
}+n_{j\downarrow }\rangle  \nonumber \\
&&+\sum_{\sigma }\{-(\langle c_{i\sigma }^{\dagger }c_{j\sigma }^{{}}\rangle
c_{j\sigma }^{\dagger }c_{i\sigma }^{{}}+h.c.)+|\langle c_{i\sigma
}^{\dagger }c_{j\sigma }^{{}}\rangle |^{2}  \nonumber \\
&&+\sum_{\sigma ^{\prime }}[(\langle c_{i\sigma }^{\dagger }c_{j\sigma
^{\prime }}^{\dagger }\rangle c_{j\sigma ^{\prime }}^{{}}c_{i\sigma
}^{{}}+h.c.)-|\langle c_{i\sigma }^{\dagger }c_{j\sigma ^{\prime }}^{\dagger
}\rangle |^{2}]\},  
\end{eqnarray}

\begin{eqnarray}
\label{de3}
c_{i\uparrow }^{\dagger }c_{j\uparrow }^{{}}(n_{i\downarrow }+n_{j\downarrow
}) &\simeq &\langle c_{i\uparrow }^{\dagger }c_{j\uparrow }^{{}}\rangle
(n_{i\downarrow }+n_{j\downarrow })+c_{i\uparrow }^{\dagger }c_{j\uparrow
}^{{}}(\langle n_{i\downarrow }\rangle +\langle n_{j\downarrow }\rangle ) 
\nonumber \\
&&-\langle c_{i\uparrow }^{\dagger }c_{j\uparrow }^{{}}\rangle (\langle
n_{i\downarrow }\rangle +\langle n_{j\downarrow }\rangle )  \nonumber \\
&&+c_{i\uparrow }^{\dagger }c_{j\downarrow }^{\dagger }\langle
c_{j\downarrow }^{{}}c_{j\uparrow }^{{}}\rangle +c_{i\uparrow }^{\dagger
}c_{i\downarrow }^{\dagger }\langle c_{i\downarrow }^{{}}c_{j\uparrow
}^{{}}\rangle  \nonumber \\
&&+\langle c_{i\uparrow }^{\dagger }c_{j\downarrow }^{\dagger }\rangle
c_{j\downarrow }^{{}}c_{j\uparrow }^{{}}+\langle c_{i\uparrow }^{\dagger
}c_{i\downarrow }^{\dagger }\rangle c_{i\downarrow }^{{}}c_{j\uparrow }^{{}}
\nonumber \\
&&-\langle c_{i\uparrow }^{\dagger }c_{j\downarrow }^{\dagger }\rangle
\langle c_{j\downarrow }^{{}}c_{j\uparrow }^{{}}\rangle -\langle
c_{i\uparrow }^{\dagger }c_{i\downarrow }^{\dagger }\rangle \langle
c_{i\downarrow }^{{}}c_{j\uparrow }^{{}}\rangle ,  
\end{eqnarray}
and the same interchanging spin up and down. Choosing one spin orientation
for the sake of clarity, the three-body terms are replaced as:

\begin{eqnarray}
\label{de4}
c_{i\uparrow }^{\dagger }c_{j\uparrow }^{{}}n_{i\downarrow }n_{j\downarrow }
&=&c_{i\uparrow }^{\dagger }c_{j\uparrow }^{{}}c_{i\downarrow }^{\dagger
}c_{i\downarrow }^{{}}c_{j\downarrow }^{\dagger }c_{j\downarrow }  \nonumber
\\
&\simeq &c_{i\uparrow }^{\dagger }c_{j\uparrow }^{{}}(\langle n_{i\downarrow
}\rangle \langle n_{j\downarrow }\rangle
+|\langle c_{i\downarrow }^{\dagger
}c_{j\downarrow }^{\dagger }\rangle |^{2}-|\langle c_{i\downarrow }^{\dagger
}c_{j\downarrow }^{{}}\rangle |^2) \nonumber \\
&&-c_{i\downarrow }^{\dagger }c_{j\downarrow }^{{}}(\langle c_{i\uparrow
}^{\dagger }c_{j\uparrow }^{{}}\rangle \langle c_{j\downarrow }^{\dagger
}c_{i\downarrow }^{{}}\rangle +\langle c_{i\uparrow }^{\dagger
}c_{j\downarrow }^{\dagger }\rangle \langle c_{i\downarrow
}^{{}}c_{j\uparrow }^{{}}\rangle )  \nonumber \\
&&-c_{j\downarrow }^{\dagger }c_{i\downarrow }^{{}}(\langle c_{i\uparrow
}^{\dagger }c_{j\uparrow }^{{}}\rangle \langle c_{i\downarrow }^{\dagger
}c_{j\downarrow }^{{}}\rangle +\langle c_{i\uparrow }^{\dagger
}c_{i\downarrow }^{\dagger }\rangle \langle c_{j\downarrow
}^{{}}c_{j\uparrow }^{{}}\rangle )  \nonumber \\
&&+n_{i\downarrow }(\langle c_{i\uparrow }^{\dagger }c_{j\uparrow
}^{{}}\rangle \langle n_{j\downarrow }\rangle +\langle c_{i\uparrow
}^{\dagger }c_{j\downarrow }^{\dagger }\rangle \langle c_{j\downarrow
}^{{}}c_{j\uparrow }^{{}}\rangle )  \nonumber \\
&&+n_{j\downarrow }(\langle c_{i\uparrow }^{\dagger }c_{j\uparrow
}^{{}}\rangle \langle n_{i\downarrow }\rangle +\langle c_{i\uparrow
}^{\dagger }c_{i\downarrow }^{\dagger }\rangle \langle c_{i\downarrow
}^{{}}c_{j\uparrow }^{{}}\rangle )  \nonumber \\
&&+c_{i\uparrow }^{\dagger }c_{i\downarrow }^{\dagger }(\langle
c_{i\downarrow }^{{}}c_{j\uparrow }^{{}}\rangle \langle n_{j\downarrow
}\rangle -\langle c_{j\downarrow }^{{}}c_{j\uparrow }^{{}}\rangle \langle
c_{j\downarrow }^{\dagger }c_{i\downarrow }^{{}}\rangle )  \nonumber \\
&&+c_{i\uparrow }^{\dagger }c_{j\downarrow }^{\dagger }(\langle
n_{i\downarrow }\rangle \langle c_{j\downarrow }^{{}}c_{j\uparrow
}^{{}}\rangle -\langle c_{i\downarrow }^{{}}c_{j\uparrow }^{{}}\rangle
\langle c_{i\downarrow }^{\dagger }c_{j\downarrow }^{{}}\rangle )  \nonumber
\\
&&+c_{i\downarrow }^{{}}c_{j\uparrow }^{{}}(\langle c_{i\uparrow }^{\dagger
}c_{i\downarrow }^{\dagger }\rangle \langle n_{j\downarrow }\rangle -\langle
c_{i\uparrow }^{\dagger }c_{j\downarrow }^{\dagger }\rangle \langle
c_{i\downarrow }^{\dagger }c_{j\downarrow }^{{}}\rangle )  \nonumber \\
&&+c_{j\downarrow }^{{}}c_{j\uparrow }^{{}}(\langle c_{i\uparrow }^{\dagger
}c_{j\downarrow }^{\dagger }\rangle \langle n_{i\downarrow }\rangle -\langle
c_{i\uparrow }^{\dagger }c_{i\downarrow }^{\dagger }\rangle \langle
c_{j\downarrow }^{\dagger }c_{i\downarrow }^{{}}\rangle )  \nonumber \\
&&+(c_{i\downarrow }^{\dagger }c_{j\downarrow }^{\dagger }\langle
c_{j\downarrow }^{{}}c_{i\downarrow }^{{}}\rangle +c_{j\downarrow
}^{{}}c_{i\downarrow }^{{}}\langle c_{i\downarrow }^{\dagger }c_{j\downarrow
}^{\dagger }\rangle )\langle c_{i\uparrow }^{\dagger }c_{j\uparrow
}^{{}}\rangle  \nonumber \\
&&+2[\langle c_{i\uparrow }^{\dagger }c_{j\uparrow }^{{}}\rangle (|\langle
c_{i\downarrow }^{\dagger }c_{j\downarrow } \rangle|^2  -
\langle n_{i\downarrow }\rangle \langle n_{j\downarrow }\rangle )  \nonumber \\
&&+\langle c_{i\uparrow }^{\dagger }c_{i\downarrow }^{\dagger }\rangle
(\langle c_{j\downarrow }^{{}}c_{j\uparrow }^{{}}\rangle \langle
c_{j\downarrow }^{\dagger }c_{i\downarrow }^{{}}\rangle -\langle
c_{i\downarrow }^{{}}c_{j\uparrow }^{{}}\rangle \langle n_{j\downarrow
}\rangle )  \nonumber \\
&&+\langle c_{i\uparrow }^{\dagger }c_{j\downarrow }^{\dagger }\rangle
(\langle c_{i\downarrow }^{{}}c_{j\uparrow }^{{}}\rangle \langle
c_{i\downarrow }^{\dagger }c_{j\downarrow }^{{}}\rangle -\langle
c_{j\downarrow }^{{}}c_{j\uparrow }^{{}}\rangle \langle n_{i\downarrow
}\rangle )  \nonumber \\
&&-\langle c_{i\uparrow }^{\dagger }c_{j\uparrow }^{{}}\rangle |\langle
c_{i\downarrow }^{\dagger }c_{j\downarrow }^{\dagger }\rangle |^{2}].
\end{eqnarray}

Each thermodynamic phase of the model in the HF-BCS approximation is
characterized by a different symmetry breaking of the expectation values
entering Eqs.~(5-8) with respect to the unperturbed
system. Since singlet \textit{s}- and \textit{d}-wave solutions do not exist
for $t_{AB}<0$ \cite{har}, we have looked for triplet \textit{p}-wave
superconductivity (TS). 
Based on symmetry properties expected for the ground state in 1D
(explained at the end of the next section), we
assumed that only the $S_z=\pm 1$ components of the triplet order
parameter do not vanish like in the Anderson-Brinkman-Model phase of
superfluid $^3He$ \cite{lege}.
For this case we have $\langle c^{\dagger}_{i\uparrow}
c^{\dagger}_{j\downarrow} \rangle $ for all $i$, $j$
in one lattice direction and
a vector $\delta $ connecting nearest-neighbours $\langle c_{i+\delta ,\sigma
}^{\dagger }c_{i,\sigma }^{\dagger }\rangle =-\langle c_{i-\delta ,\sigma }^{\dagger
}c_{i,\sigma }^{\dagger }\rangle =\psi \neq 0$, while in other directions $%
\langle c_{i+\delta ,\sigma }^{\dagger }c_{i,\sigma }^{\dagger
}\rangle =0$. As a
consequence also $\langle c_{i+\delta ,\sigma }^{\dagger }c_{i,\sigma
}\rangle $ depends
on direction. We also considered the usual CDW and SDW phases for which the
cubic lattice in D dimensions is divided into two equal interpenetrating
sublattices A and B, in such a way that the nearest-neighbours of any site
of A lie in B. In this case, for the SDW $\langle n_{i,\sigma }\rangle
=(1+m\sigma e^{i%
\mathbf{Q.R}_{i}})/2$ while for the CDW $\langle n_{i,\sigma }\rangle =(1+\Delta e^{i%
\mathbf{Q.R}_{i}})/2$ with $\vec{Q}=(\pi ,\pi ,...,\pi )$ and $m$ and $%
\Delta $ order parameters. The resulting one-particle Hamiltonian has the
form of the non-interacting one with a renormalized effective hopping $%
t_{eff}$, plus a symmetry breaking perturbation which, at half filling,
depends on two or three parameters to be determined selfconsistently. These
parameters are the corresponding order parameter ($\psi $, $m$ or $\Delta $)
and the different values of $\langle c_{i+\delta ,\sigma }^{\dagger
}c_{i,\sigma }\rangle $: one for CDW, SDW, or TS in 1D, two for TS in
more than 1D (for TS in more than 1D $t_{eff}$ becomes anisotropic). 

The phase diagram has a natural energy scale which we call $E$: the absolute
value of the energy of the half-filled non-interacting case for hopping $%
|t_{3}|=t_{AA}+t_{BB}-2t_{AB}$, counting both spins. The ratio $E/|t_{3}|$
depends only on the dimension of the simple cubic lattice and is $4/\pi
=1.273$ in 1D, $16/\pi ^{2}=1.621$ in 2D, 2.005 in 3D \cite{yoko} and 0.798
in $\infty $D \cite{yoko}. The only point in the $U-V$ phase diagram for
which the paramagnetic solution is stable is the triple point $%
(U_{t},V_{t})=(E,E/z)$ where $z=2D$ is the coordination number. If and only
if $V<V_{t}$, for any $U$, the paramagnetic solution is unstable against the
TS phase; if and only if $U>U_{t}=E$, it is unstable against SDW; and if and
only if $V>(E+U)/2z$ it is unstable against the CDW. These boundaries allow
us to establish that, in any dimension, inside the region bounded by the
dashed line in Fig. 1, the TS is the stable phase. Also the CDW-TS boundary
lies between the dashed and dot-dashed lines of Fig. 1. With an adequate
change of variables, it can be shown that the self-consistency problems for
the CDW and SDW take the same form on the line $U=zV$. Thus, in any
dimension, for $U \geq E$, $U=zV$ is the boundary of the CDW-SDW first-order
transition, independently of the other parameters of the model.

The above results are quite general and valid in any dimension. The region
of stability of the TS extends always beyond the dashed lines of Fig. 1 and
the precise location of the boundaries of the TS  depends on $t_{AB}$
and the dimension. In Fig. 1, the specific
case of 1D and $t_{AA}=t_{BB}=1$, $t_{AB}=0.2$ is shown. The change in slope
of the SDW-TS boundary is due to a metamagnetic first-order transition
inside the SDW from large to small order parameter as $V$ increases. A
similar situation occurs in the CDW-TS boundary. Near the triple point, and
particularly for $t_{AB}\sim 1$, all order parameters become very small, and
the numerical method used to solve the self-consistency equations breaks
down. Similarly, in 1D, for $V=0$ and $t_{AB}=0.097$, the effective hopping
vanishes in the paramagnetic phase and the HF-BCS approximation becomes
invalid for $t_{AB}\sim 0.1$ or smaller.

One expects that the HF-BCS results are reliable for small values of the
interactions (small $U$, $V$ and $1-t_{AB}$, with $t_{AA}=t_{BB}=1$) and D$%
>1 $, supporting the existence of the \textit{p}-wave TS for these
parameters. In 1D there is no true LRO and the HF-BCS results only have
qualitative validity. A comparison with more accurate methods in 1D is made
in the next section.

\section{Phase diagram obtained from topological transitions}

In recent years, the concept of Berry phase was a subject of great interest
in a variety of fields in physics. Zak has shown that it can be used for
labeling energy band in solids \cite{zak}, and subsequent work showed that
changes of polarization are proportional to the corresponding change in a
Berry phase \cite{rest}. Ortiz and Martin have generalized these concepts to
a many-body ground state \cite{orti}, and Resta and Sorella used this
concept to identify a ferroelectric transition \cite{re2}. This many-body
Berry phase is simply the phase captured by the ground state in a ring of $L$
sites as the boundary conditions $c_{i+L,\sigma }^{\dagger }=e^{i\phi }$ 
$c_{i,\sigma }^{\dagger }$, complete a cycle from a flux $\phi =0$ to $\phi
=2\pi $ \cite{berry,orti,re2,or2}. We call this phase the charge Berry phase 
$\gamma _{c}$.

While previous work assumed always that the ground state is non-degenerate
for all $\phi $ (except at isolated points of the parameter space at which $%
\gamma _{c}$ is indefinite), two of us have recently generalized this
concept for the case in which there is a crossing of levels in the ground
state as a function of $\phi $ \cite{berry}. This is the case of the
superconducting phases, for which there is AFQ as a consequence of the
crossing energy levels. Furthermore, a ``spin'' Berry phase $\gamma _{s}$
was defined as that captured by the ground state in the cycle $0\le \phi \le
2\pi $ varying the boundary conditions as $c_{i+L,\sigma }^{\dagger }=e
^{i\sigma \phi }c_{i,\sigma }^{\dagger }$ with $\sigma =1$ (-1) for spin up
(down) \cite{berry}. Due to the inversion symmetry of the model Eq.~(\ref
{eqn:3}). the Berry phases are quantized and in 1D can take only the values
0 and $\pi $ (modulo $2\pi $). Thus, it turns out that $\gamma _{c}/\pi $
and $\gamma _{s}/\pi $ are topological quantum numbers which are related to
changes in the total polarization and the difference between polarizations
for spin up and down respectively \cite{berry}.

Defining a vector $\gamma=(\gamma_c,\gamma_s)$, it is easy to see \cite
{berry} that for a CDW (SDW) with maximum order parameter one has $\gamma=(0,0)$ ($%
\gamma=(\pi,\pi)$), and these values are consistent with the change in up
and down polarizations when the electrons of spin down of one phase are
moved to the other sublattice to form the other phase. Since $\gamma$ jumps
in $\pi$ only at the phase transitions, the values $\gamma$(CDW)=(0,0) and $%
\gamma$(SDW)=$(\pi,\pi)$ are valid for any non-zero magnitude of the
corresponding order parameters. These values are also consistent with the
canonical transformation.

\begin{equation}  \label{eqn:8}
c^{\prime}_{j,\uparrow}=c_{j,\uparrow}, c^{\prime}_{j,\downarrow}=(-1)^j
c^{\dagger}_{j,\downarrow}
\end{equation}
under which the CDW and SDW are interchanged and the Berry phases modulo $%
\pi $ transform as:

\begin{equation}  \label{eqn:9}
\gamma^{\prime}=\gamma+\pi
\end{equation}

Since the ground state for $U=V=0$ is invariant under Eq.~({\ref{eqn:8}),
Eq.~({\ref{eqn:9}) implies that for the third phase present in the diagram $%
\gamma _{c}(TS)=\gamma _{s}(TS)+\pi $ (numerically it turns out that $\gamma
(TS)=(0,\pi )$), and thus, at least one of the topological numbers $\gamma
_{c}/\pi $, $\gamma _{s}/\pi $ jumps at each boundary. Thus, this method
combined with finite-size scaling is able to determine accurately the phase
diagram (see Fig.~2). }}

The best previous numerical methods to determine phase diagrams of this type
were based on the size dependence of SDW and CDW order parameters, and their
probability densities in a Monte Carlo sampling \cite{mon,hir2,cann}.
However, these quantities as well as different correlation functions vary
smoothly at the transition and it is very difficult to obtain accurate
boundaries \cite{hir2,cann2,cann,four}. 
Instead, the use of topological numbers as order parameters
necessarily leads to sharp transitions.

In Fig.~2 we show the phase diagram of the model Eq.~(\ref{eqn:3}) in 1D
obtained with the above mentioned method, for two values of $t_{AB}$, and
compare it with the corresponding HF BCS results and with the exact result
for $t_{AB}\to 0$ \cite{14}. Although, as mentioned in the previous section,
the HF BCS results are not expected to have quantitative validity in 1D for
large values of the interaction, for small $t_{AB}$ the resulting phase
diagram is in reasonable agreement with that obtained with the above
explained more reliable numerical method. The results of the latter tend to
the exact phase diagram as $t_{AB}\to 0$ . The comparison also shows that
the TS phase in 1D extends beyond the expectations of the HF BCS results.
This is particularly clear for $t_{AB}\sim t_{AA}=t_{BB}$ and $U\sim 2V$,
and is probably related to the fact that, in the continuum limit theory, the
backscattering and Umklapp terms coming from the $U$ and $V$ terms of the Hamiltonian
(ultimately responsible of the insulating behaviour), nearly cancel each
other on the line $U=2V$ \cite{cann,voit}.

We should note that the method of the topological transitions alone, is not
able to identify the nature of the TS phase, since different phases can have
the same topological numbers. For example, we find $\gamma=(0,0)$ not only
for the CDW, but also in the negative-$U$ Hubbard model with small negative $%
V$, for which \textit{s}-wave singlet pairing occurs \cite{voit,lin}. We
also find it for our model (Eq.~(\ref{eqn:3})) with $t_3=V=0$ and small $U$,
for which also singlet superconducting correlations dominate at large
distances, as explained in Section 2.

In the following we summarize the evidence in favour of the triplet
superconducting phase in 1D.

a) \textit{Superconductivity}:

\begin{itemize}
\item  Since each time the Berry phase is evaluated, the ground state energy
as a function of the flux $E(\phi )$ is computed at the same time, we have
checked that there is a tendency to AFQ (i.e. $E(\phi )\simeq E(\phi +\pi )$%
) inside almost all the TS phase.

\item  Our previous calculations of the correlation exponent $K_{\rho }$ for 
$V=0$ show the dominance of superconducting correlations at large distances (%
$K_{\rho }>1$) for $U<U_{c}$, and an opening of a charge gap for $U>U_{c}$ 
\cite{mit,13}. The value of $U_{c}$ has been estimated as $U_{c}\sim 3.5$
for $t_{AB}=0.2$ and $t_{AA}=t_{BB}=1$ \cite{13} in good agreement with the
result shown in Fig.~2. For $t_{AB}=0.6$, from $K_{\rho }$ one obtains $%
U_{c}=2.05\pm 0.05$ \cite{mit}, while from the topological transition we
obtain $U_{c}=2.11$ \cite{berry}.

\item  For a ``metallic'' gapless phase with $K_{\rho }\le 1$ (as in the
non-interacting case), one expects undetermined Berry phases \cite{berry,or2}
instead of the result $\gamma =(0,\pi )$ for the TS phase.
\end{itemize}

b) \textit{Triplet character}:

\begin{itemize}
\item  The HF BCS decoupling in any dimension leads to unstable singlet and
stable triplet superconductivity.

\item  For known cases of singlet superconductivity (mentioned above) we
obtain $\gamma =(0,0)$ in contrast to the result $\gamma (TS)=(0,\pi )$.

\item  We find that in rings of 10 sites the nearest-neighbour
triplet-triplet correlations are larger than the singlet-singlet ones at the
largest distance in the ring (5 sites).

\item  Numerically, we obtain a non-degenerate ground state inside the TS
phase which should transform into itself under the symmetry transformation (%
\ref{eqn:8}) for $U=V=0$, since the Hamiltonian is invariant at that point.
An ordinary BCS singlet solution transforms into a SDW under this
transformation, leading necessarily to a degenerate ground state for $U=V=0$%
. Instead, using $c^{\prime }{}_{k,\downarrow }^{\dagger
}=c_{-k+Q,\downarrow }$, $|0\rangle =\prod_{k}c^{\prime }{}_{k,\downarrow
}^{\dagger }$ and $\cos (k_{\alpha }+Q_{\alpha })=-\cos (k_{\alpha })$, it
can be easily checked that our {\it p}-wave triplet BCS  solution in the $x$%
-direction (see Section 3 after Eq.~(8)): 
\begin{equation}
|p_{x}>=\prod_{k\sigma ,k_{x}>0}(u_{k}+v_{k}c_{k,\sigma }^{\dagger
}c_{-k,\sigma }^{\dagger })|0\rangle
\end{equation}
is invariant under Eq.~(\ref{eqn:8}).


\item  Finite-size scaling is consistent with the absence of a spin gap
inside the TS phase. Continuum-limit theory predicts that triplet
superconducting correlations dominate at large distances (due to logarithmic
corrections) when $K_{\rho }>1$ and a spin gap is absent \cite{voi2}.
\end{itemize}

\section{Summary and discussion}

We have studied the occurrence of superconductivity in the generalized
Hubbard model Eq.~(\ref{eqn:3}). This model contains the most important
physical ingredients expected to describe transition metals in general \cite
{voll} or cuprate superconductors \cite{marsiglio,simon,fedro,den}. In 1D,
most of the parameters of the model appropriate for systems like
trans-polyacetylene have been estimated \cite{camp,pain} and are consistent
with several regions of parameters for which we find superconductivity.

When the three-body term $t_{3}=2t_{AB}-t_{AA}-t_{BB}$ vanishes, singlet 
\textit{s}-wave superconductivity is expected for electron densities per
site $n\neq 1$ or very small values of $U$. For $n$ not too near one and $%
V=0 $, Eq.~(\ref{eqn:4}) describes qualitatively the values of $U$ below
which superconducting correlations dominate at large distances in 1D.
However, numerical results suggest that Eq.~(\ref{eqn:4}) underestimates
these values by a factor $\sim 1.4$. For $n=1$, the region of
superconducting behaviour is very small.

Instead, for $n=1$ and $t_{AB}<t_{AA}=t_{BB}=1$, we find evidence of
a \textit{p}-wave triplet superconducting phase (TS) for small values of $U$
and $V$. The Hartree-Fock BCS approximation gives a stable TS for
sufficiently small $U$ and $V$ in any dimension. In 1D, the numerical method
of the Berry phases, described in the previous section, predicts a stable TS
even for small $1-t_{AB}$ and large values of $U\sim 3$ near the line $U=2V$.

While a definite proof of the triplet character of this phase does not exist
so far, there are several arguments in favour if it enumerated at the end of the previous
section. In contrast to the previous case of \textit{s}-wave
superconductivity ($2t_{AB}=t_{AA}+t_{BB}$) mentioned above, for $%
t_{AB}<t_{AA}=t_{BB}$, the effect of doping seems to weaken the TS in favour
of the SDW at $V=0$ \cite{13}.

Experimentally, evidence of a triplet odd-parity superconducting order
parameter exists in some quasi-1D organic conductors \cite{bebe}. In
1D, our solution is compatible with a nodeless superconducting gap, as
evidenced by thermal conductivity measurements in $(TMTSF)_2ClO_4$ \cite{bebe}.

The effect of phonons, particularly in 1D, can stabilize insulating
CDW states, or favour $s$-wave superconducting states. However, based
on the general study performed in 1D using bosonization \cite{voi2},
we believe that our results remain qualitatively valid in presence of a
small or moderate electron-phonon coupling. In addition, if phonons
are treated in the antiadiabatic approximation, the model Eq. (3)
retains its form [21b].

\section{Acknowledgments}

Two of us (A.~A.~A. and E.~G.) would like to thank Prof. J. Voit for useful
discussions concerning the nature of the \textit{p}-wave superconducting
phase and Prof. G. Ortiz for discussions about topological quantum numbers.
E.~G. and K.~H. are supported by Consejo Nacional de Investigaciones
Cient\'{\i}ficas y T\'ecnicas (CONICET). A.~A.~A. is partially supported by
CONICET.

\newpage\ 

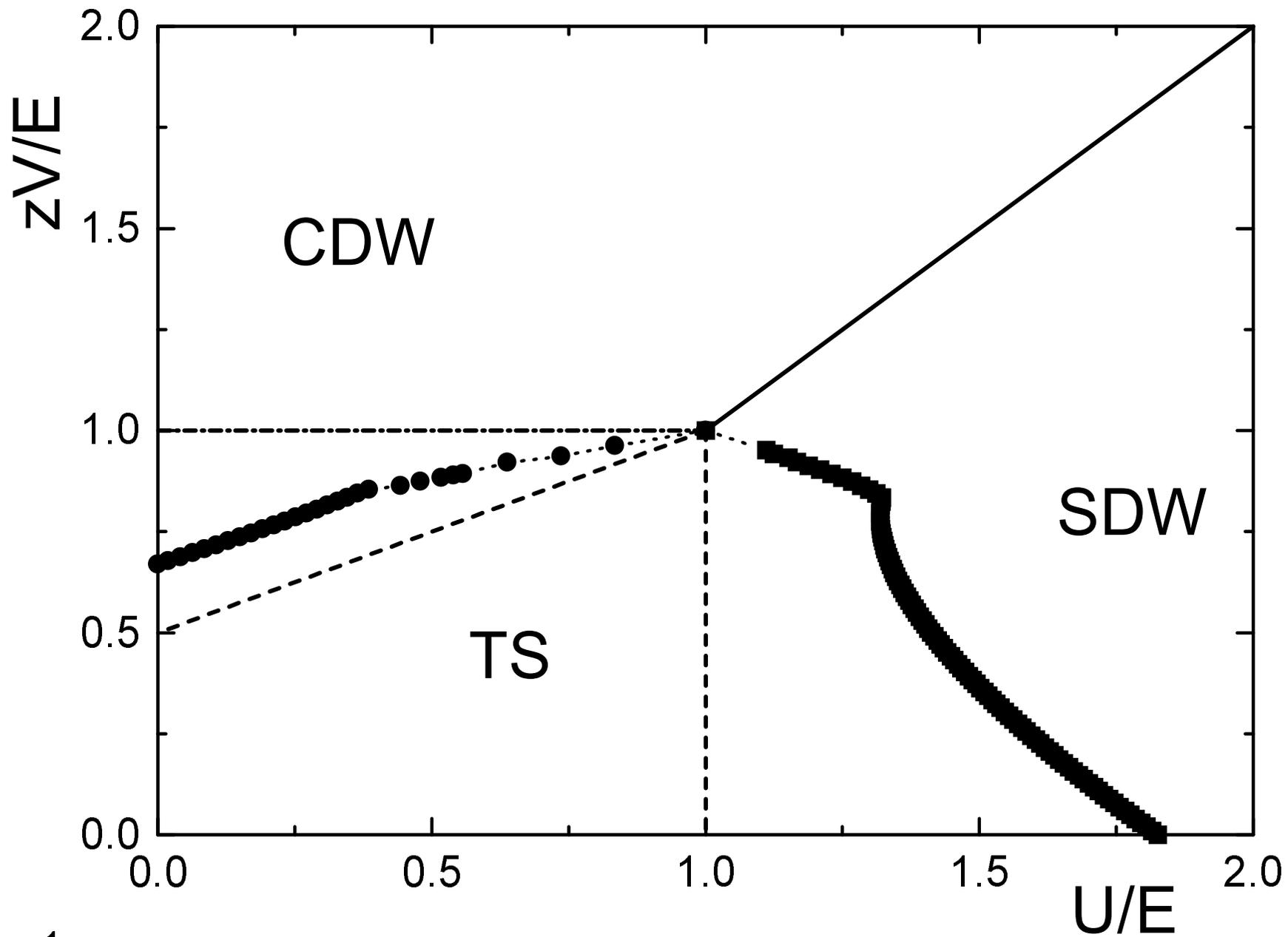
Figure 1: Mean-field (HF BCS) phase diagram of the model (\ref{eqn:3}) for
one particle per site and $t_{AA}=t_{BB}>t_{AB}$. The full line is the
CDW-SDW boundary in any dimension. Inside the region bounded by the dashed
lines, the triplet superconducting (TS) phase is the stable one. The CDW-TS
boundary lies between the dashed and dot-dashed lines. $E/2(t_{AA}-t_{AB})$
is a number of order one which depends on dimension (see text). Solid
circles (squares) are points on the CDW-TS (SDW-TS) boundary in 1D for $%
t_{AA}=t_{BB}=1$ and $t_{AB}=0.2$.

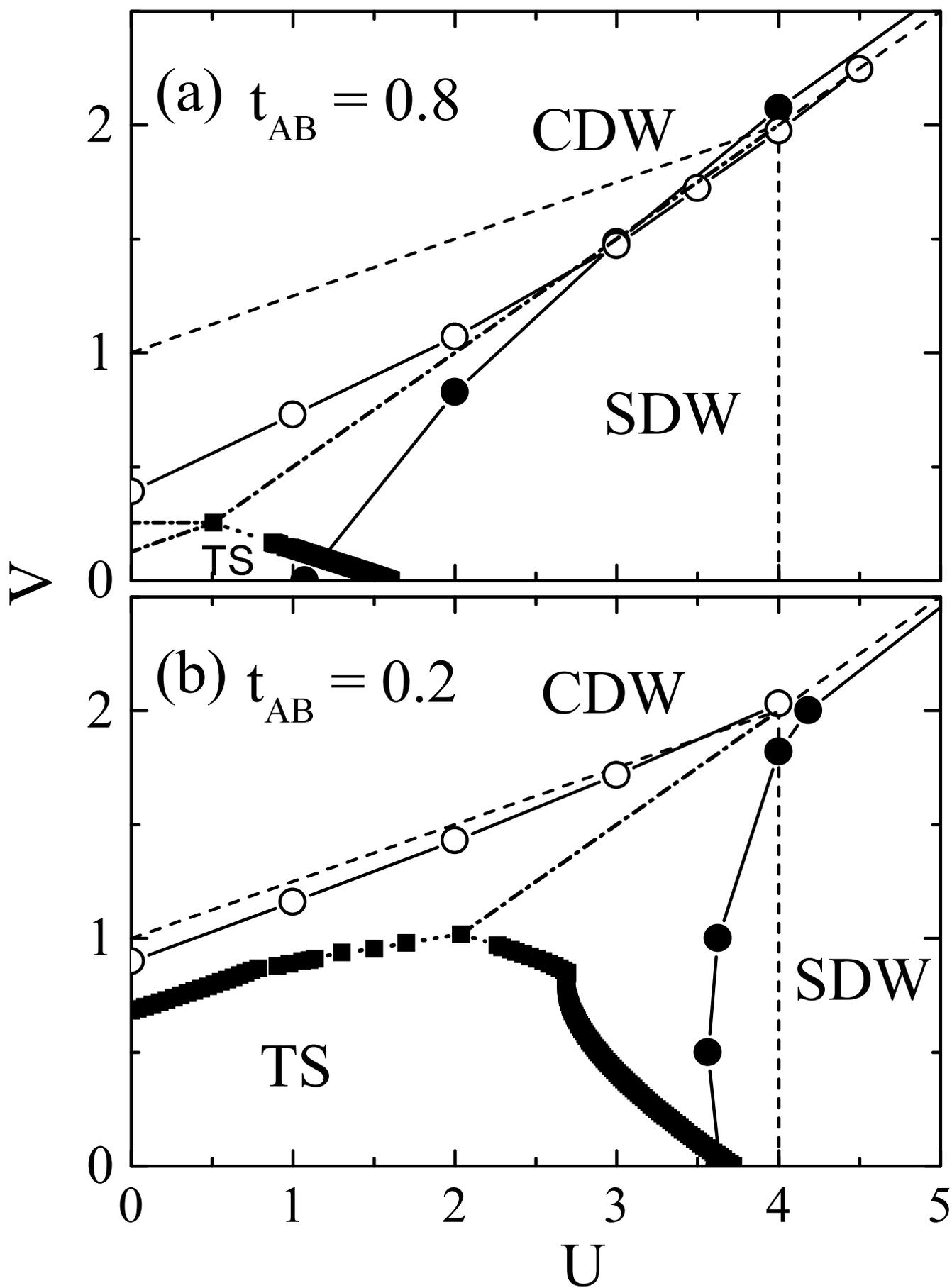
Figure 2: Phase diagram of the model (\ref{eqn:3}) in 1D for one particle
per site, $t_{AA}=t_{BB}=1$ and two different values of $t_{AB}$. Solid
(open) circles are determined from the jump of $\gamma_c$ ($\gamma_s$).
Solid squares, dotted and dot-dashed lines are HF BCS results. The dashed
lines are exact results in the limit $t_{AB}\to 0$ \protect\cite{14}. Full
lines are guides to the eye. The small difference between the results from
both Berry phases at the CDW-SDW boundary is due to finite-size
effects.

\end{document}